# An active feedback recovery technique from disruption events induced by m=2 n=1 tearing modes in ohmically heated tokamak plasmas

*P. Zanca, R. Paccagnella, C. Finotti, A. Fassina, G. Manduchi, R. Cavazzana, P. Franz, C. Piron, L. Piron*

*Consorzio RFX (CNR, ENEA, INFN, Università di Padova, Acciaierie Venete Spa) Corso Stati Unit 4, 35127 Padova (Italy)*

**Abstract**

We present experimental results of magnetic feedback control on the $m=2$, $n=1$ tearing mode in RFX-mod operated as a circular ohmically heated tokamak. The feedback suppression of the non-resonant $m=2$, $n=1$ Resistive Wall Mode (RWM) in $q(a)<2$ plasmas is a well-established result of RFX-mod. The control of the tearing counterpart, which develops in $q(a)>2$ equilibrium, is instead a more difficult issue. In fact, the disruption induced by a growing amplitude $m=2$, $n=1$ tearing mode can be prevented by feedback only when the resonant surface $q=2$ is close to the plasma edge, namely $2<q(a)<2.5$, and the electron density does not exceed approximately half of the Greenwald limit. A combined technique of tearing mode and $q(a)$ control has been therefore developed to recover the discharge from the most critical conditions: the potentially disruptive tearing mode is converted into the relatively benign RWM by suddenly decreasing $q(a)$ below 2. The experiments demonstrate the concept with 100% of successful cases. The $q(a)$ control has been performed through the plasma current, given the capability of the toroidal loop-voltage power supply of RFX-mod. We also propose a path for controlling $q(a)$ by acting on the plasma shape, which could be applied to medium size elongated tokamaks.

**1. Introduction:**

Disruption events have been detected in tokamaks since the very beginning of the experimental campaigns on these plasmas. A comprehensive description of this phenomenon and the variety of factors which can trigger is provided in [1, 2]. Empirically detected limits for relatively safe operation in terms of plasma pressure normalized to the magnetic field (beta), plasma density, and safety factor at the plasma edge *q(a)* have become universally recognized textbook concepts [3]. They are respectively the Troyon beta-limit [4], the Greenwald density limit [5], and the *q(a)*>2 constraint. The latter prevents the destabilization of the *m*=2, *n*=1 non-resonant ideal external kink [3], converted into a Resistive Wall Mode (RWM) by a shell close enough to the plasma [6, 7]. In recent years the introduction of electro-magnetic feedback performed by active coils has demonstrated the possibility of stabilizing RWM [8]. A significant contribution in this field has been given by RFX-mod operated as a circular ohmic tokamak with on-axis toroidal field of 0.55T [9]. In fact, non-disruptive plasmas with *q(a)*<2 have been realized thanks to the feedback suppression of the *m*=2, *n*=1 RWM, by means of active coils placed outside the passive conductive structures (i.e. vacuum-vessel and stabilizing shell) [10, 11]. The control has proven to be successful even using only six active coils in the outboard mid-plane [12]. The result is by itself quite important, since it extends the tokamak operation to a region of parameters considered forbidden in the past, which could be extrapolated to reactor relevant performances by a new high-field compact device [13]. Apart from the above mentioned limits, a major cause of disruption for standard plasma with *q(a)*>2 is wall-locking of the *m*=2, *n*=1 tearing mode [3], the resonant counterpart of the previous RWM. Magnetic feedback control of wall-locked tearing modes is a well-assessed technique in the reversed field pinch (RFP) operations of RFX-mod [9, 14]. The same kind of control on the *m*=2, *n*=1 tearing mode has been recently experimented in the RFX-mod *q(a)*>2 tokamak discharges, at the same time of similar operations realized by DIII-D in high-beta shaped plasma using in-vessel coils [15]. These experiments deal with slowly rotating tearing modes and show significant analogies despite the different layout and plasma configuration. We have also to mention previous feedback experiments on high frequency *m*=2, *n*=1 tearing modes realized in DITE [16] and HBT-EP [17] by in-vessel coils having large bandwidth power supplies. The present paper discusses the RFX-mod results. Differently from its RWM counterpart, we have found that a tearing mode cannot be suppressed by the RFX-mod feedback, but only mitigated in its saturation amplitude. Moreover, the feedback against a growing amplitude *m*=2, *n*=1 tearing mode has proven to be sufficient to avoid the disruption only in a low-*q(a)* region, 2<*q(a)*<2.5, and in conditions of moderate plasma density. On the basis of the successful and rather straightforward RWM control, this paper presents the alternative idea of stabilizing the $m=2$, $n=1$ tearing mode by converting it into the non-resonant counterpart. This involves an equilibrium control simultaneous to the magnetic feedback, which drives $q(a)$ below 2 as soon as the potentially disruptive tearing mode is detected. In order to demonstrate the concept, we have exploited the capabilities of the RFX-mod toroidal loop-voltage power supply system [18], which can be used to increase quickly the plasma current and therefore reduce $q(a)$. In RFX-mod this is a safe operation since tokamak discharges are realized at a current level well below the machine limits. The experiments demonstrate the 100% success rate of this technique in recovering the plasma from disruptive condition, using the $q(a)<2$ equilibrium as escape route. These results encourage the exploration of an alternative technique, based on the shape control, to access the $q(a)<2$ equilibrium starting from a standard configuration. The paper content is divided as follows. In section 2 we provide a general description of the magnetic feedback on $m=2$, $n=1$ tearing mode in RFX-mod. Some model based interpretations are given in section 3. The statistic of the $m=2$, $n=1$ induced disruptions is presented in section 4 with a classification of the mode rotation frequency regimes. In section 5 the results of the new technique of simultaneous $q(a)$ and mode control are described. Finally, a discussion about the applicability of the $q(a)$ control by acting on the plasma shape is given in the concluding section 6. A description of the RFXlocking code, used for simulations of the $m=2$, $n=1$ tearing mode feedback is given in appendix A. In appendix B two approximate methods are described for the estimate of the on-axis $q(0)$, necessary quantity in modeling the plasma equilibrium.

**2. General considerations on $m=2$, $n=1$ tearing mode and its magnetic feedback**

The RFX-mod plasma ($a=0.459$m, $R_0=2$m) is contained by a small time-constant ($\tau_v \approx 3$ms) inconel vacuum vessel (average radius $r_v=0.49$m) and passively stabilized by a more external 3mm thick copper shell ($\tau_w=100$ms, inner radius $r_{wi}=0.5125$m). By managing in real-time a large amount of signals [19, 20], RFX-mod is able to accomplish simultaneously feedback on magneto-hydro-dinamic (MHD) instabilities and equilibrium control. The latter basically involves plasma current, through powerful solid-state power supplies sized for the high toroidal voltage requests of the RFP configuration [18], and plasma shape, through poloidal field coils. The MHD control is realized by a grid of 4(poloidal)×48(toroidal) active saddle loops placed on the outermost support structure

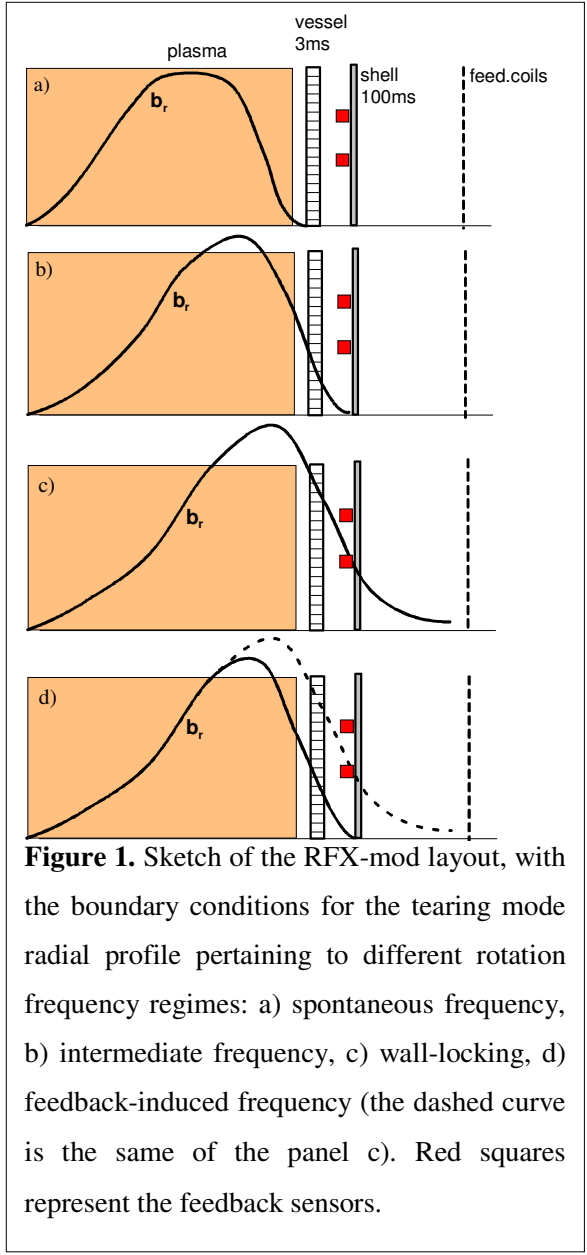

**Figure 1.** Sketch of the RFX-mod layout, with the boundary conditions for the tearing mode radial profile pertaining to different rotation frequency regimes: a) spontaneous frequency, b) intermediate frequency, c) wall-locking, d) feedback-induced frequency (the dashed curve is the same of the panel c). Red squares represent the feedback sensors.

($\tau_s \approx$ 24ms) at radius $c$=0.5815m [21]. The magnetic sensors used by MHD feedback are placed close to the shell inner surface at radius $r_s$=0.507m. Due to the screening of the vacuum-vessel these sensors cannot detect high frequency MHD phenomena. To this purpose, and only for off-line analyses, another set of large band-width probes located on the vessel inner surface (ISIS system) [22] is used. The $m$=2, $n$=1 tearing mode is generally present in the RFX-mod tokamak discharges. When this mode rotates at its fast natural frequency, which amounts to some kHz, it can be seen only by the ISIS sensors, since the vessel acts as a perfectly conducting shell in this frequency range. Equilibrium modifications can increase the mode amplitude and the ensuing stronger interaction with the eddy currents induced onto the vessel slows down the rotation, so the mode progressively penetrates the vessel and becomes visible also on the feedback sensors. As shown by the more quantitative analysis discussed later in section 4, an intermediate rotation regime with frequencies of several hundred Hertz is detected by these sensors. At these frequencies the copper shell provides an almost ideal boundary, which screens both the mode and any feedback field produced by the active coils. Therefore, feedback is ineffective on this rotation branch. Without feedback a complete stop (wall-locking) occurs if an amplitude threshold at the resonant surface is exceeded [23, 24]: the amplitude further grows as a consequence of the penetration of the shell and a disruption invariably occurs. At this stage feedback control comes into play. In fact, by keeping small the edge amplitude it pushes the otherwise wall-locked mode into rotation with angular frequencies significantly smaller than the natural ones (and also smaller than the intermediate ones in RFX-mod) yet much larger than the inverse of the shell time-constant. Consequently, the shell penetration is hindered and the saturation amplitude of the mode mitigated [25]. The feedback-induced frequencies represent a dynamical equilibrium in the momentum equation, which establishes when the feedback gain is above a critical value [14, 15]. If $\Delta t$ represents the global delay of the feedback chain, the angular frequency is of order $\omega \sim 1/\Delta t$ [15]. In general $\omega\tau_w \gg 1$, so the mode amplitude saturates under the constraint of a quasi-ideal boundary close to the shell location. In the presence of several passive structures between plasma and active coils, the previous statement remains valid considering an 'effective shell', given by a sort of barycenter of the structures weighed by the relative time constant [26]. In RFX-mod copper shell plays the major role during the feedback action. This 'wall-unlocking' by feedback allows RFP operation up to 2MA in RFX-mod. In figure 1 a qualitative representation is given of the mode boundary conditions in the cases above discussed. Figure 2 shows a typical example of the feedback action on a $m=2$, $n=1$ tearing mode, which tends to wall-lock. The considered shot has $q(a) \approx 2.1$, according to the large aspect-ratio circular cross-section estimate $q(a) = 2\pi a^2 B_{0\phi}/(\mu_0 I_p R_0)$, being $B_{0\phi}$ the toroidal magnetic field and $I_p$ the plasma current [3]. Plot a) shows the non-integrated radial field signal detected by one of the ISIS sensors (black) and the integrated radial field signal measured by one of the feedback sensors (red). The feedback-induced rotation, at about 60-70Hz in this case, appears as soon as the mode slows down to become visible on the feedback sensors. Plot b) displays the final part of the discharge, with the black line representing the plasma current. As soon as the feedback is deactivated at 0.8s, the mode wall-locks and a disruption occurs.

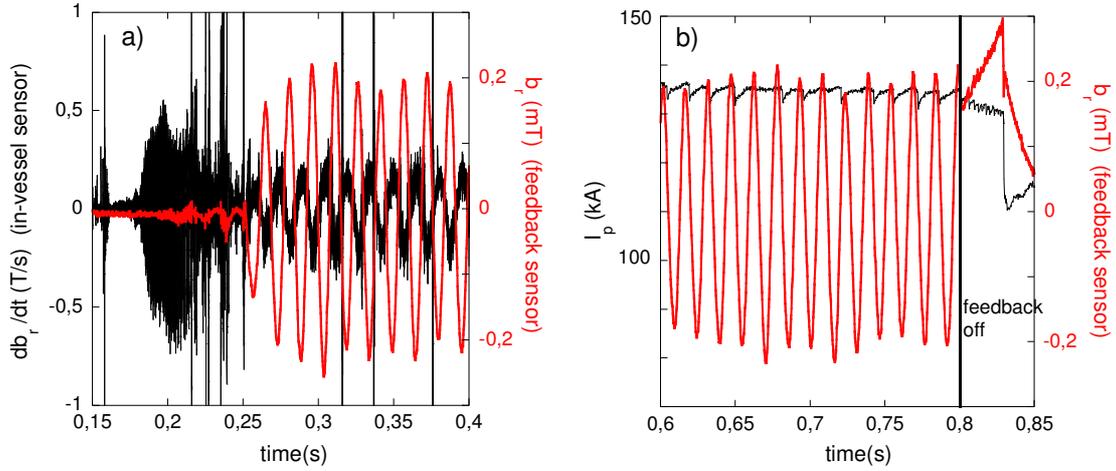

**Figure 2.** Feedback control on the *m*=2, *n*=1 tearing mode in the shot 33748. In red we show the feedback radial field integrated signal for one sensor, in the first part of the discharge (a) and in the second part (b). In black we have the ISIS non-integrated radial signal for one sensor (a) and the plasma current (b).

In RFX-mod the key elements to obtain the feedback-induced rotations is the removal in the feedback variable of the aliasing generated by the active coils' sideband harmonics. This is implemented in real-time by the Clean-Mode-Control technique (CMC) [14]. The example shown in figure 2 applies CMC to the radial field sensors' signal. Once such aliasing is minimized, the experimental feedback results are little dependent on the acquired magnetic field component (radial and poloidal fields have been used to control the *m*=2, *n*=1 tearing so far), apart from the different intervals of effective gains. Even the plasma-sensors distance should be unimportant, if the latter are placed inside the most conductive wall. This is predicted by simulations discussed in the next section. Moreover, it is experimentally confirmed by the equivalence, as feedback variables, between the radial sensors' signal and its extrapolation at the plasma radius computed with the inclusion of the perturbed toroidal field signal. The ultimate reason is in fact that the feedback-produced boundary condition is mainly determined by the conductive walls layout between plasma and coils.

However, the avoidance of wall-locking and shell penetration realized by feedback is a necessary but not a sufficient condition to prevent the *m*=2, *n*=1 tearing induced disruption in RFX-mod. In many cases the radial field diffusion across the vessel and the slowing down of the mode to the intermediate frequency regime is sufficient to trigger a disruption. The most critical conditions are discriminated by the equilibrium parameter *q(a)* and by the plasma density. This issue will be addressed in section 4.

## 3. Simulation of the *m*=2, *n*=1 feedback control

The RFXlocking cylindrical code [25, 27] adapted to the tokamak configuration simulates rather well the interaction of the *m*=2, *n*=1 tearing mode with the external conductive structures including feedback in RFX-mod. A detailed description of the code is postponed to appendix A. The model computes the mode frequency and saturated amplitude, but cannot predict the disruption occurrence. The equilibrium is fixed by a zero-pressure, Wesson-like current profile $\mu_0 \mathbf{J}_0 = \sigma(r)\mathbf{B}_0$, $\sigma(r) = \sigma_0 \cdot \left(1-(r/a)^2\right)^{\alpha}$ [3], which should be adequate for the low-*β*, ohmic, circular plasma of RFX-mod. The two parameters of this model are related to the on-axis and edge values of *q* by $\sigma_0 = 2/(q(0)R_0)$, $\alpha \approx q(a)/q(0) - 1$. Figure 3 present simulations for equilibrium where wall-locking is predicted without feedback. Mode amplitude at the resonant surface, i.e. island width, is kept fixed at a very low level up to *time*=0.06s, which is the assumed viscous diffusion time, in order to give the plasma flow the possibility to relax to the equilibrium profile. Afterwards, the island width evolves with the Rutherford equation [28]. In the simulation without feedback we note two steep frequency drops: the first corresponds to the radial field diffusion through the vessel; the second, leading the rotation to negligible values, is the wall-locking and involves the penetration of the shell. Instead, in the simulation with feedback the latter drop is avoided and the frequency is maintained at about 60Hz. The island width saturates at a value determined by the quasi-ideal boundary realized by the feedback close to the shell. Here, CMC on radial field is considered, assuming the sensors just on the shell inner surface ($r_s = r_{wi}$).

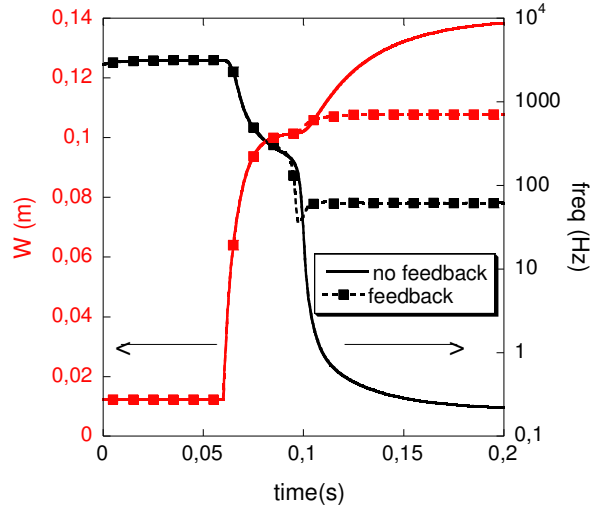

**Figure 3.** RFXlocking $m=2$, $n=1$ simulations for the equilibrium $q(a)=2.68$, $q(0)=1$. Continuous and dashed lines with symbols refer to non-controlled and feedback simulations respectively. In red we plot the island width and in black its rotation frequency. Note the log-scale for the frequency.

The equivalence between radial sensors located at different radii, mentioned in the previous section, is demonstrated in figure 4 by the good superposition of the frequency and island width curves when plotted as function of the coils' current $m=2$, $n=1$ harmonic amplitude. This equivalence also highlights the feedback ultimate limit in the mode amplitude reduction, which is fixed by the passive conductive boundary between plasma and coils. When active coils are placed inside the vacuum-vessel, control delays and amplifiers finite power become the most important limitations for the feedback, as suggested by the experiments [16, 17] and the theoretical analysis [27].

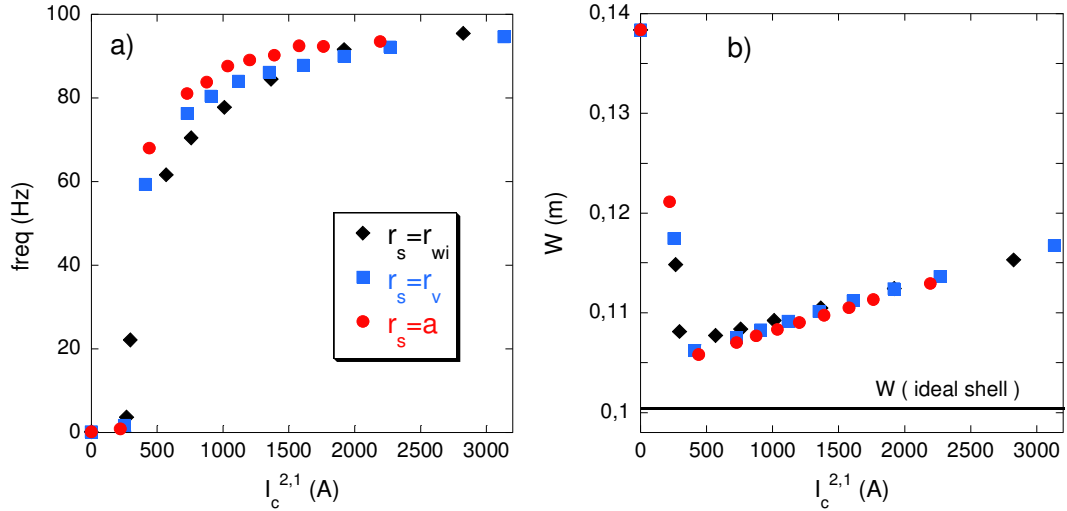

**Figure 4.** RFXlocking $m=2$, $n=1$ simulations considering CMC applied to the radial field at different radii: the shell inner surface (black diamonds), the vacuum vessel average radius (blue squares), and the plasma radius (red circles). The different points are obtained varying the feedback gains, with the same equilibrium considered in figure 3. Rotation frequency (a) and island width (b) at the final saturated amplitude state are plotted. The saturated island width in the presence of an ideal shell in the place of the copper one, for the given equilibrium, is also shown in plot b) (black line). The amplitude of the control current harmonic is considered in the x-axis.

## 4. Statistic of the disruptions induced by the $m=2$, $n=1$ tearing mode in RFX-mod

Experiments and simulations agree on the fact that a magnetic feedback with coils placed outside the main shell cannot in general suppress the $m=2$, $n=1$ tearing mode, unlike its RWM counterpart, but only keep controlled its saturation amplitude. In RFX-mod there are equilibrium conditions in which this is not sufficient to avoid a disruption. For a proper discussion of this topic, we first present in a more quantitative way the rotation frequency regimes introduced in section 2. Figure 5 plots the experimental mode frequency detected by the feedback sensors for different $q(a)$ (colored points), both with and without feedback. The color code refers to the disruption statistic and it will be discussed later in this section. RFX-mod has little explored the $q(a)>3.5$ region, since the relatively low toroidal field requires very small plasma current to access it. The black and grey diamonds refer to the steady state solution of RFXlocking simulations (with and w/o feedback respectively) performed under different equilibria. As shown in appendix B an approximate estimate of the on-axis $q(0)$ can be given on the basis of the core temperature and the edge magnetic measurements, upon some simplifying assumptions. It is found that $q(0)$ increases with $q(a)$, as shown by figure 14 in appendix B. The general trend of this figure is used to perform the equilibrium scan in simulation. The experimental natural frequency branch is not visible on the feedback sensor due to the vessel screening, so it is not reported in the figure. The three slower frequency branches, defined in section 2 (intermediate, feedback-induced, wall-locked), order the experimental points, and are also recovered in simulation. The wall-locked, approximately below 10Hz, is mostly obtained without control in the low-$q(a)$ interval $2<q(a)<2.5$. In fact, the interaction between the mode and the first-wall becomes stronger, increasing the likelihood of wall-locking, as the resonant $q=2$ surface moves towards the plasma edge. The feedback branch is the previous one increased in the range 10-100Hz by the control action: the dispersion of the experimental points is explained by the variety of CMC techniques and gains gathered here, whereas the simulated black points refer only to CMC on radial field with a fixed gain. The intermediate branch is distributed in the entire interval considered for $q(a)$. Many non-controlled cases (blue circles) mix with the controlled ones (red squares and green triangles) within this branch, to confirm that this frequency range is little affected by feedback. In simulation the intermediate branch is the lower region (below 1 kHz roughly speaking) of the upper band of solutions, which extends close to the natural values not detectable by the feedback sensors in the experiment. As clearly shown by the color code of figure 5, the correlation between wall-locking and disruptions, and therefore the possibility to prevent the latter by the feedback action, can be established only at low $q(a)$ ($2<q(a)<2.5$). In general, larger $q(a)$ can cause disruptions without wall-locking. More precisely, we have no evidence of these events when the mode is rotating in the natural branch. But, once the mode slows down diffusing across the vacuum-vessel and becoming visible on the feedback sensors, at $q(a)>2.5$ there is no way to avoid the disruption, though the mode frequency is mostly in the intermediate branch. Instead, for $2<q(a)<2.5$, if wall-locking is removed by feedback, disruptions are much less frequent and caused in general by high density, which is the other important parameter as we shall see later (this is the case for all the red squares with $q(a)<2.5$ but one, in figure 5).

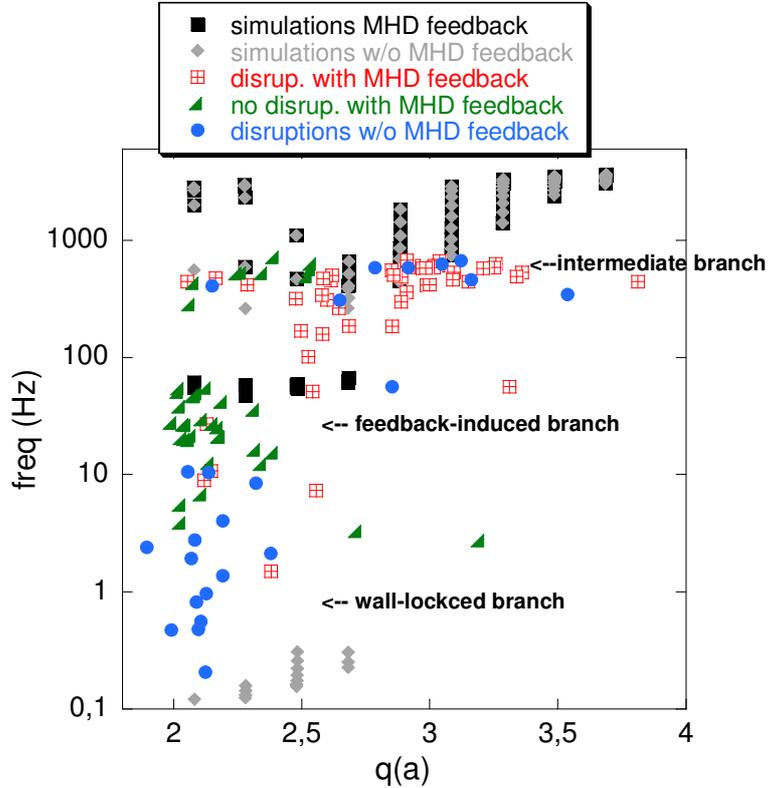

**Figure 5.** Rotation frequencies of the $m=2$, $n=1$ tearing mode, as detected by the feedback sensors, plotted against $q(a)$. Coloured symbols are experimental values, averaged over 5ms before the disruption for the blue circles (shots without feedback) and red squares (shots with feedback), and 5ms around the maximum detected mode amplitude for the green triangles, which correspond to the not disrupted shots with feedback. Black and grey symbols are simulated frequencies with and without feedback respectively, considering different equilibria.

The $q(a)$ dependence of the disruption incidence for the feedback controlled shots cannot be explained by the radial field amplitude at the edge, which seems to decrease with $q(a)$ (figure 6a), rather by the estimated island width $W$, which is instead found to increase with $q(a)$. This reflects the similar correlation, but in this case decreasing, between $W$ and the estimated radius $r_{mn}$ of the $q=2$ surface. A possible interpretation is that the saturation amplitude of the mode increases with the distance of this surface from the stabilizing shell. In such a case, the plasma-shell distance fixed by the machine layout would determine an offset in this trend. In figure 6b) the portion of the island which stands inside the plasma, $W_{eff}=min(W/2,r_{mn})+min(W/2,a-r_{mn})$, is plotted against $q(a)$: $W_{eff}$ should be in fact the relevant quantity for the plasma confinement. The radial field amplitude at the resonant surface $|b_r(r_{mn})|$ used for computing $W$, according to the standard island-width formula reported in Appendix A, is the extrapolation of the edge data given by the Newcomb's equation solution over the Wesson-like equilibrium model (actually the cylindrical version of the method described in [29]). In a particular shot the latter is fixed by $q(a)$, as given by the external magnetics, and by the estimate of the on-axis $q(0)$ from the internal inductance $l_i$, exploiting the method described in appendix B (see figure 14 there). This Newcomb's extrapolation suffers for neglecting the screening of the vacuum-vessel, so for the intermediate frequencies cases it may be an under-estimate of the true island width. The computed island width can exceed half plasma radius.

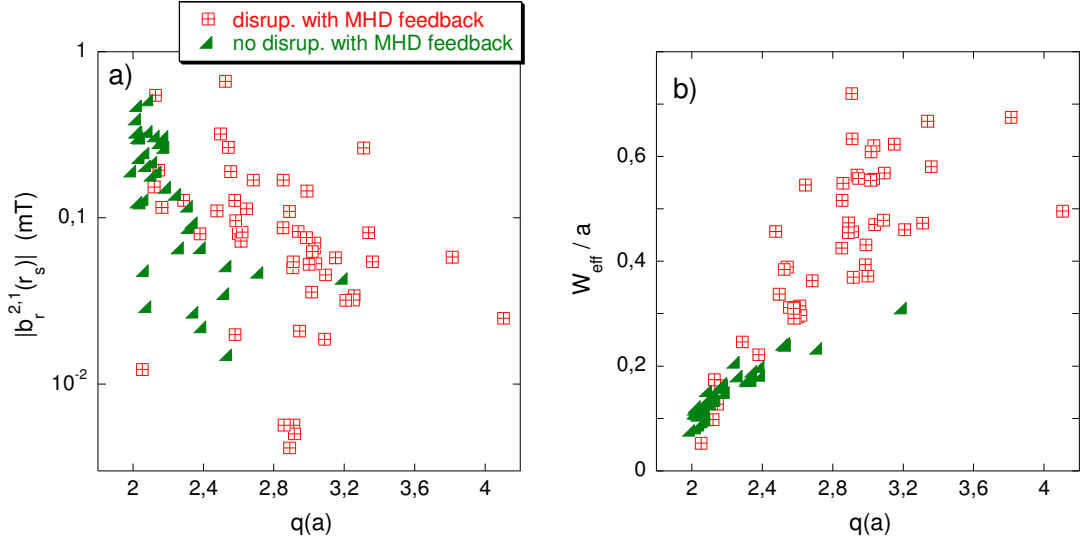

**Figure 6.** Dependence on $q(a)$ of the $m=2$, $n=1$ mode amplitude for the feedback shots considered in figure 5 (red disrupted, green not disrupted). The amplitudes refer to the maximum detected values (in a 5ms time interval before the disruption for red squares). Plot a): radial field at the sensors' radius. Plot b): island width inside the plasma normalized to the plasma radius.

Though most of the disrupted cases correspond to larger values of $W_{eff}$, the partial superposition with the non-disrupted points in figure 6b) suggests the existence of another important parameter. This is the plasma density: as shown in figure 7, the two sets of points separate well when $q(a)$ or $W_{eff}$ are plotted against the Greenwald electron density fraction $n_e/n_G$, being $n_G=I_p(MA)/(\pi a^2)\times 10^{20}$ [5]. Figure 7a) shows that the non-disruptive region is restricted to $2 \leq q(a) \leq 2.5$ and moderate plasma density, $n_e \leq 0.5 n_G$. Figure 7b) can be interpreted in terms of a density limit which decreases, eventually going to zero, with the island width, or equivalently in terms of a critical island width which decreases, eventually going to zero, with the density. These criticalities should depend on the plasma transport properties, but not on the machine passive layout. For this reason we believe that the trend shown by figure 7b) is quite fundamental. Instead, the pattern of figure 7a) should inherit the offset due to the plasma-shell distance from the relationship between $W_{eff}$ and $q(a)$. It is conceivable that in a circular device with a better plasma-shell proximity the non-disruptive $q(a)$ region should be larger than in RFX-mod. Note that the red and green points are separated by a nearly empty region. This may be due to an insufficient statistic. However, it could also indicate that for some combinations of $n_e/n_G$ and $q(a)$ the $m=2$, $n=1$ tearing mode, always present in the RFX-mod discharges, is unlikely to abandon the natural rotation frequency branch where it remains invisible for the feedback sensors used in this analysis. Further investigations will try to clarify this point. Past DITE observations, about the increase of the disruptive density limit when applying feedback on the $m=2$, $n=1$ tearing mode, in particular at $q(a)$ close to 2 [16], are in consonance with the RFX-mod results.

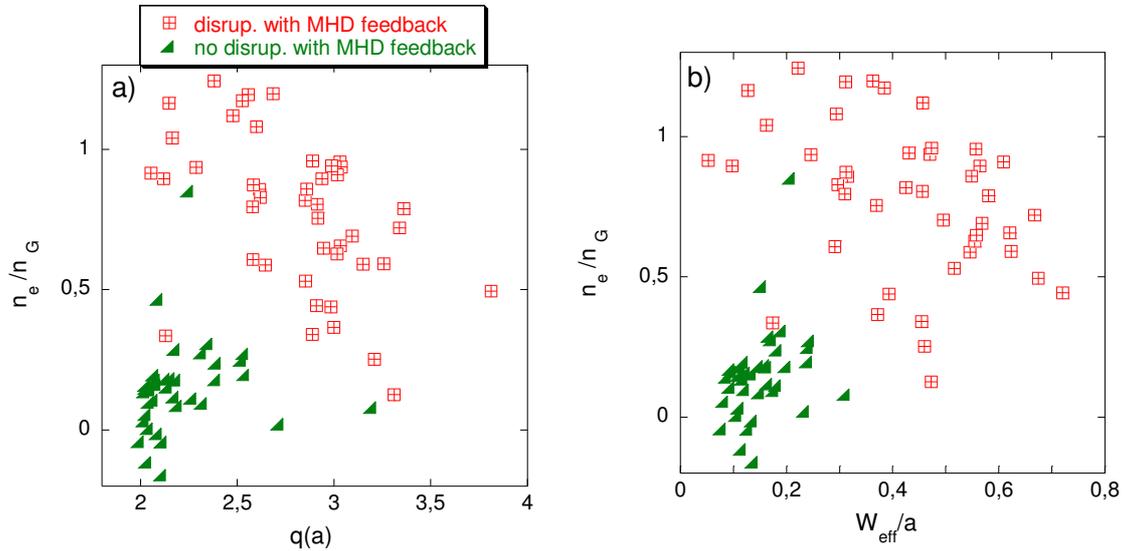

**Figure 7.** Line-average electron density from interferometer data [30] normalized to the Greenwald limit for the feedback shots considered in figures 5, 6 (red disrupted, green not disrupted) plotted against $q(a)$ (a) and the normalized island width (b). A negative density means a very small, badly measured value.

In conclusion, we ascribe the ultimate reason of the $m=2$, $n=1$ tearing induced disruption to the presence of a large magnetic island and possibly to the transport enhancement it implies, rather than to the wall-locking of the mode itself. The island is particularly detrimental at high density, perhaps due to a coupling with the radiation losses (a possible interpretation of this effect is given in [31]). Moreover, we point out the limitation of the magnetic feedback, which avoids wall-locking and the ensuing perturbation diffusion across the shell, but cannot prevent the penetration of the innermost vacuum-vessel, which appears to be critical in many conditions.

## 5. Simultaneous control of the $m=2$, $n=1$ tearing mode and $q(a)$

The previous analysis leads quite naturally to the idea that disruption prevention should include the control of the parameter $q(a)$ in order to ensure favorable equilibrium conditions for doing feedback on the $m=2$, $n=1$ mode. In the radical solution we propose, the detection of the $m=2$, $n=1$ tearing mode on the feedback sensors triggers the equilibrium control and ramps down $q(a)$ below 2, thereby converting the tearing mode into a pure RWM. Then, feedback is able to suppress this more benign mode, and plasmas with density near to the Greenwald limit can be sustained. The purpose is to demonstrate the simple concept that the removal of the $m=2$, $n=1$ resonant surface, and consequently of the magnetic island, restores a good non-disruptive discharge, though with characteristics different from the initial one.

Before going into the details of this scheme, let's briefly describe the phenomenology of the $m=2$, $n=1$ induced disruptions in RFX-mod. An example is given in figure 8a): a growing amplitude $m=2$, $n=1$ tearing mode (blue line) involves a sudden crash to nearly zero-level of the soft-X-rays signal (SXR) from tomography [32] (red line), which is generally due to a drop both in the temperature and in the density, as we will see later. The concomitant positive spike in the plasma current (black line), associated to a similar negative spike in the toroidal voltage (not shown), is ascribed by [33] to the current profile flattening implied by the disruptive magnetic island. With the power supplies available in standard tokamaks the confinement loss caused by an event of this kind makes impossible to sustain the configuration, and a current quench occurs. In RFX-mod the current quench is induced within a soft-stop scheme, which interrupts the current feedback control, bringing to zero the applied toroidal voltage, after the detection of the first spike. On the contrary, i.e. without interruption of the current sustainment, the power supplies capability would allow maintaining the initial current level, producing a series of disruptions following the very first one (figure 8b). This means that the discharge cannot be recovered if we try to keep the equilibrium fixed, because the $q=2$ resonant surface, where the disruptive island can grow, continues to be present inside the plasma.

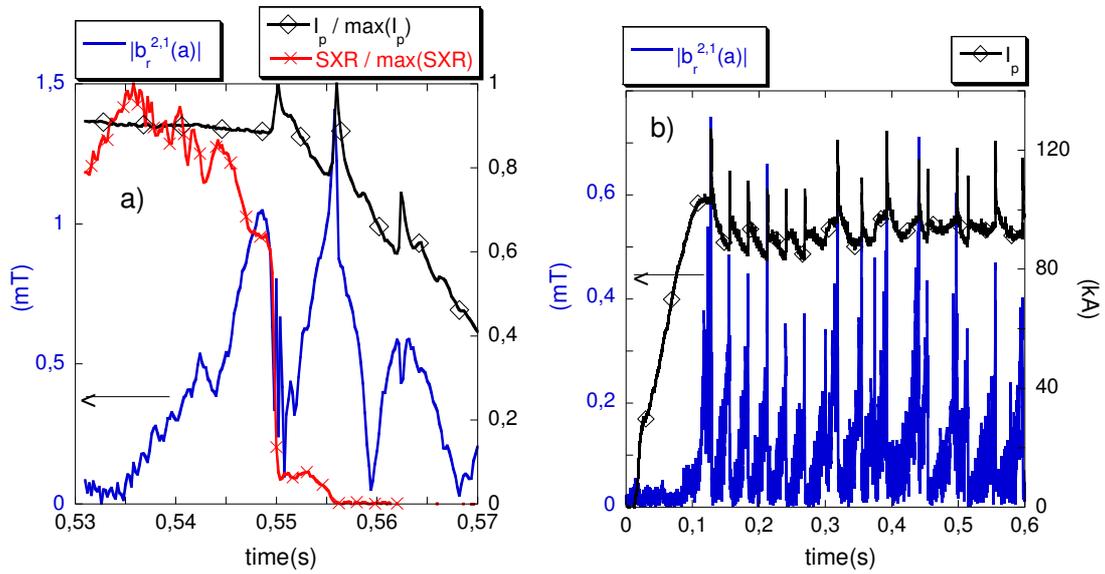

**Figure 8.** a): in correspondence of a disruption event at about 0.55s for the shot 33594 ($q(a)=2.5$), we plot the radial field amplitude extrapolated at the plasma surface (blue), the normalized (to the maximum) soft-X-ray emissivity (red line with crosses), and the normalized plasma current (black line with diamonds); the applied toroidal voltage is switched off after the first positive current spike at about 0.55 and then the current decays. b): for the shot 36347 ($q(a)=3$) we plot the radial field amplitude at the plasma surface (blue) and the plasma current (black line with diamonds); a series of disruptions, marked by the current spikes, occur because the current feedback control tries to maintain the $q(a)=3$ magnetic equilibrium.

The Thomson scattering diagnostic (TS) [34] provides the electron temperature and density with a good deal of spatial resolution. However, we have to underline the difficulties of such measurements in these low current plasmas. In figure 9 the core values are plotted for some disrupted shots with MHD feedback (those with reliable TS data) around the SXR crash time. The disruption implies a drop in both signals. Nonetheless, in RFX-mod the temperature does not collapse to few eV, as often reported for other devices [33]. The points well after the drop are not of interest, since they correspond to the plasma current induced decay.

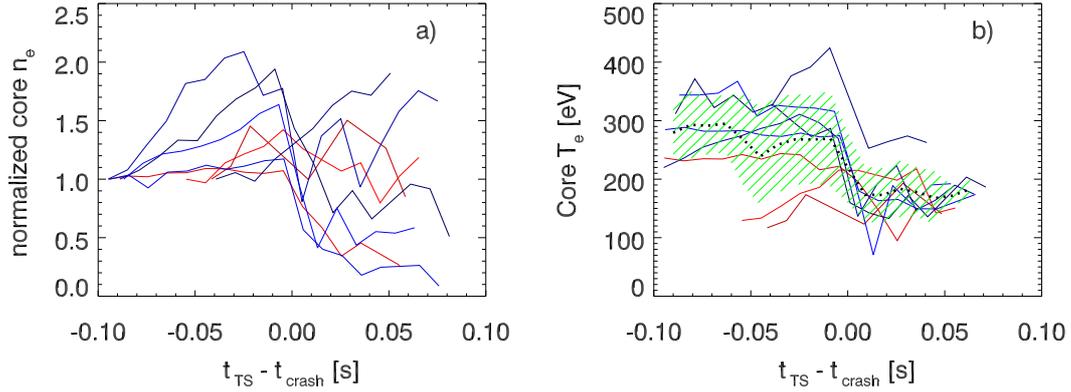

**Figure 9.** Electron core density normalized to the first plotted value (a), and electron core temperature (b), as measured by TS. A time interval centered about the SXR crash caused by the disruption is considered. The TS sampling time is 10ms. Different lines correspond to different disrupted shots. Those with a more evident temperature drop are plotted in shades of blue. In plot b) the dotted thick line and band are respectively the shot-average temperature and its standard deviation.

The scheme we propose tries instead to save the discharge coupling the MHD control to the $q(a)$ control, which removes the resonant surface necessary for the island to exist. In RFX-mod the simplest way to control $q(a)$ is by acting on the plasma current: in fact the very flexible and powerful toroidal loop voltage circuit allows a rapid increase of $I_p$ such to reduce $q(a)$ below 2 from an initial value above 3 in about 10÷15ms. A series of experiments have then been performed following this sequence: 1) a tokamak with $q(a)$ around 3 is initially programmed; 2) feedback control on the $m=2$, $n=1$ tearing mode is applied keeping at the same time monitored its amplitude in the poloidal field (the largest perturbed component); 3) when this amplitude exceeds a threshold, set to 0.6mT on the basis of a statistical analysis of the disruptions, a positive pulse of toroidal loop voltage, shaped according to a pre-programmed waveform, produces a rapid increase of the current against the plasma inductance, leading $q(a)$ below 2; 4) a new steady-state $q(a)<2$ tokamak is formed, where $q(a)$ is maintained close to the reference value by the current control, feedback stabilized against the $m=2$, $n=1$ mode, now a pure RWM; on average, the final toroidal voltage is slightly larger than the initial value. A typical example is shown in figure 10. Starting from an initial equilibrium with $q(a)\approx3$ (plot a), the profiles modification induced by the approaching of plasma density to the Greenwald limit (plot b) destabilizes the $m=2$, $n=1$ tearing mode. As soon as its amplitude exceeds the threshold, the $q(a)$ control intervenes with a positive $V_{loop}$ pulse (plot c). The tearing mode is converted into a RWM, suppressed during the equilibrium transition phase (plot d), and a stable $q(a)<2$ discharge with $n_e/n_G\approx0.6$ (plots a, b) settles down at the end. Looking at plot d) we note that the disruptive tearing mode (the first one) disappears during the ramp down when $q(a)$ is still larger than 2. According to the statistic of figure 7a), the second tearing mode, destabilized when $q(a)$ is close to 2 and seeding the following RWM, is of a more benign type. This technique has been 100% successful so far without false alarms. In about half of the experiments the temperature and density drop produced by the mode triggering the $q(a)$ control could not be avoided. This is shown by the statistical analysis of the TS data plotted in figure 11: the drops, when occur, are similar to those of the standard disruptions of figure 9, but they are not systematic. The shot considered in figure 10 is one of these cases, characterized by a confinement loss. Nonetheless, a significant thermal energy content in the $q(a)<2$ phase is restored, as can be inferred from plot 10b). This is a consequence of the magnetic island removal, which re-establishes the confinement to acceptable level, rather than of the additional ohmic heating brought by the current increase. In fact the total degradation of the confinement in the presence of a disruptive island would nullify any attempt of maintaining the global thermal energy, as demonstrated by the sequence of events in the shot plotted in figure 8b). Figures 12, 13 illustrate some further statistical analyses. Using figure 7a) as background, figure 12 displays the plasma conditions, in terms of $q(a)$ and Greenwald density fraction, which this technique allows to recover from (black squares): most of them would inevitably lead to a disruptive plasma. The final plasma equilibrium, which settles down after the $q(a)$ ramp, is also shown (blue squares): the points compare rather well with the standard, stationary $q(a)<2$ shots (green diamonds).

Figure 13 compares the standard $q(a)<2$ shots with the same equilibria after the $q(a)$ ramp down, from the thermal energy point of view. According to the TS core electron temperature, and the poloidal beta estimated from the diamagnetic flux (see appendix B), $q(a)<2$ discharges similar to the standard ones establish upon magnetic island removal.

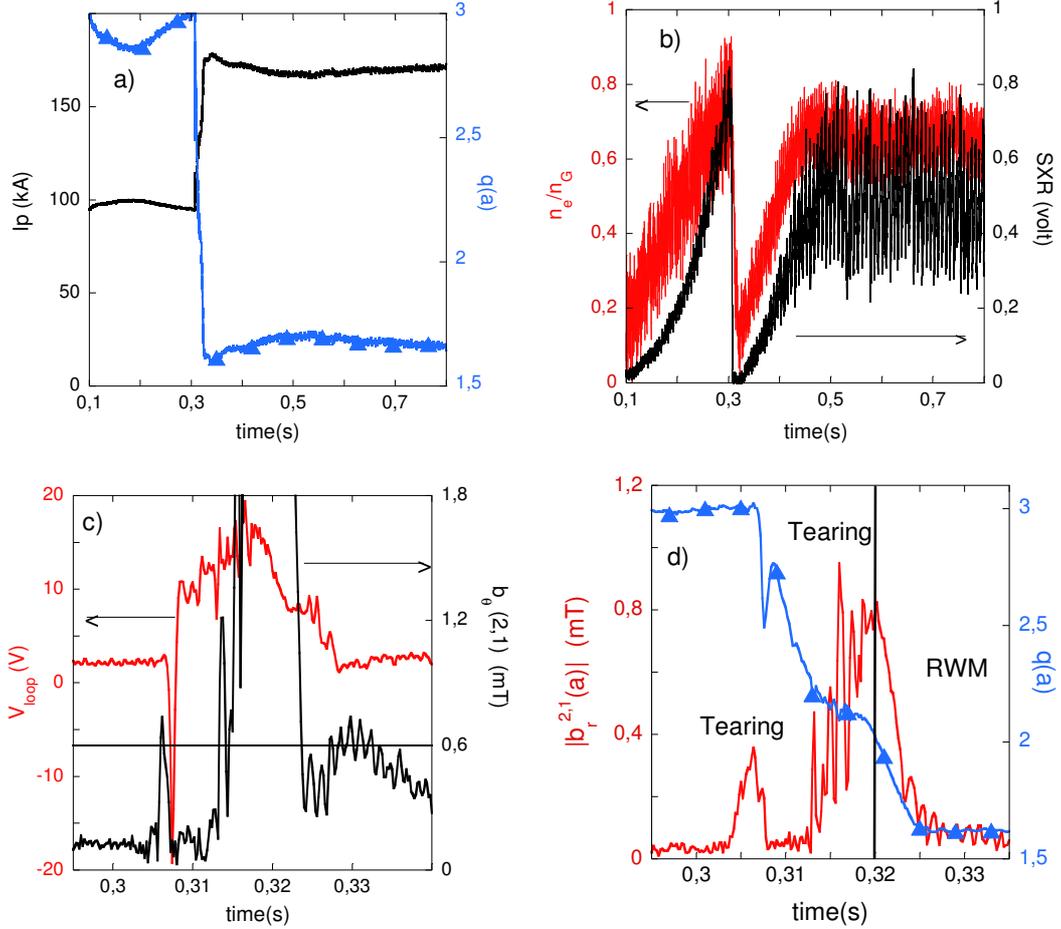

**Figure 10.** Waveforms for the shot 36359 with simultaneous $m=2$, $n=1$ and $q(a)$ control. a): plasma current (black) and $q(a)$ (blue with symbols) from external magnetics. b): line-average electron density from interferometer normalized to the Greenwald limit (red), and line-integrated SXR signal (black). c): toroidal voltage (red), and monitored mode amplitude in the poloidal field component (black; the straight thick line is the threshold). d) amplitude of the radial field extrapolated at the plasma surface (red), and $q(a)$ (blue with symbols) during the equilibrium transition phase.

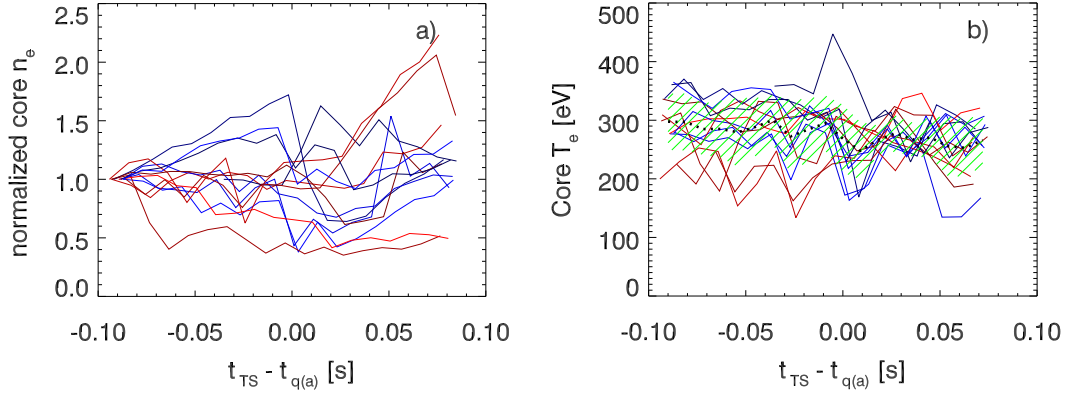

**Figure 11.** Electron core density normalized to the first plotted value (a), and electron core temperature (b), as measured by TS for the discharges with simultaneous MHD and *q(a)* control. A time interval is considered around the triggering of the *q(a)* control . The TS sampling time is 10ms. Different lines correspond to different shots. Those with a more evident temperature drop are plotted in shades of blue. In plot b) the dotted thick line and band are respectively the shot-average temperature and its standard deviation.

The amplitude threshold, which triggers the equilibrium modification, has been prudentially set to a relatively large value to minimize the risk of false alarms produced by spurious effects. In general the growth of the disruptive instability is very fast (few ms), and it is not preceded by any clear precursor (the slow growing case in figure 8a is an exception). Due to experimental time limitation we did not have the possibility of testing smaller threshold values. Future experiments will tell if the sometimes observed initial confinement loss event could be avoided by a more severe threshold, which activates earlier the *q(a)* ramp down. With respect to the relatively cold discharges of RFX-mod, we expect that the physical constraints are more relaxed in a large tokamak, due to the higher temperature and the consequent longer relaxation times. Nonetheless, in this case technological limitations might be an issue. The final section will speculate on this topic. At the present status, we can conclude that our technique is a well-established recovery method from potentially-disruptive or even disrupted conditions.

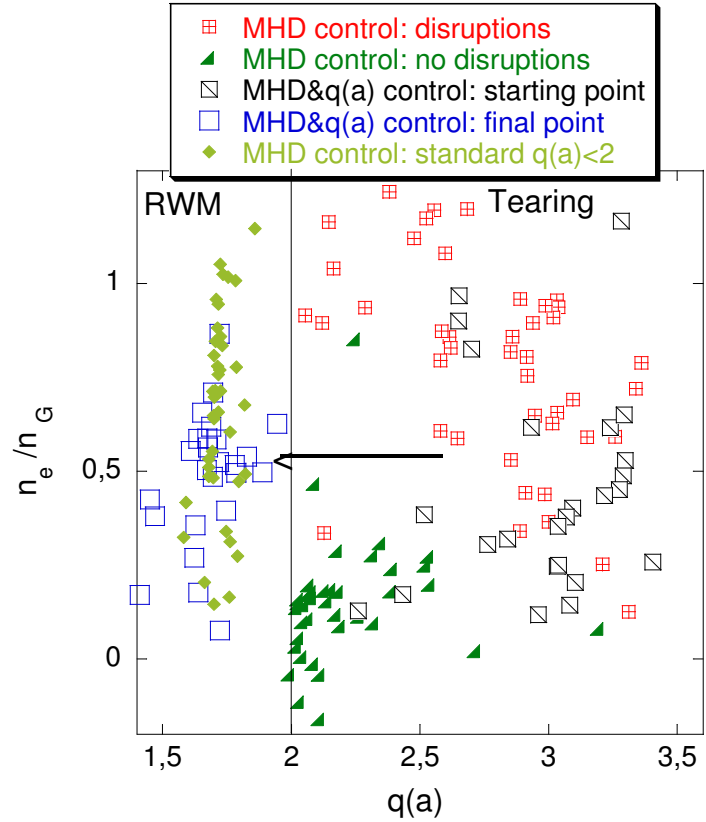

**Figure 12.** Summary of the feedback operation against the $m=2$, $n=1$ mode in terms of Greenwald density fraction (line-averaged data from the interferometer) and $q(a)$. Pure MHD control on the tearing mode for green triangles and red squares (same points of figure 7a). Pure MHD control on the RWM for the standard $q(a)<2$ shots represented by green diamonds (data averaged in a 50ms time interval). The other symbols are the simultaneous MHD and $q(a)$ control data: black squares denote the application of the $q(a)$ ramp-down (data averaged in a 4ms time interval in correspondence of the potentially disruptive $m=2$, $n=1$ tearing mode) and blue squares are the final equilibrium at $q(a)<2$ (data averaged in a 50ms time interval starting 100ms after the application of the current ramp).

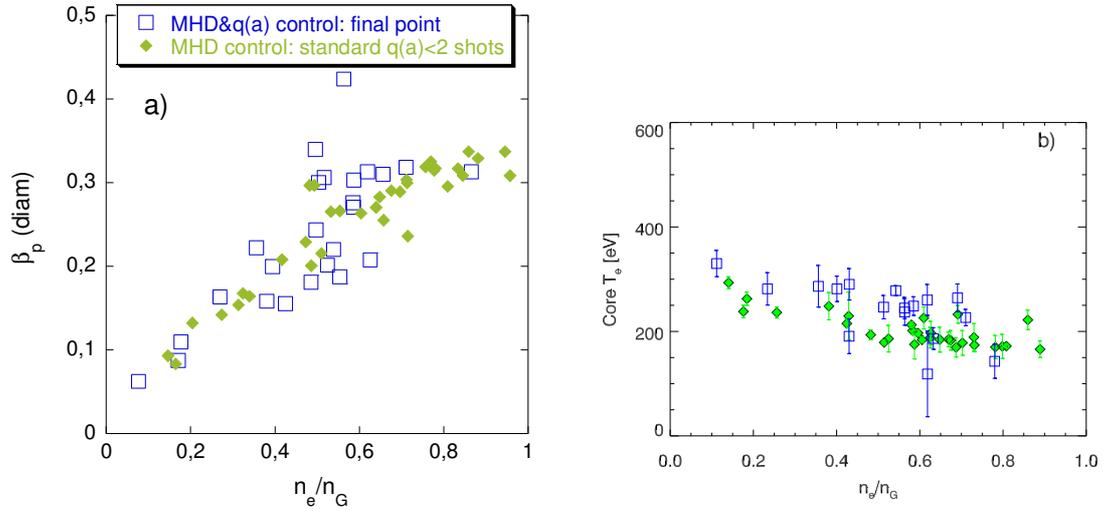

**Figure 13.** Comparison between standard $q(a)<2$ equilibrium (green diamonds) and final state of the $q(a)$ control (blue squares). a): poloidal beta from diamagnetic flux plotted against the Greenwald electron density fraction from interferometer (line averaged data). b): electron core temperature plotted against the Greenwald electron density fraction from TS data (error bars included).

## 6. Conclusions and future perspectives.

In RFX-mod, operated as a circular Ohmic tokamak, feedback control of the $m=2$, $n=1$ tearing mode with active saddle coils placed outside the shell has proven to be a difficult task. Differently from its RWM counterpart emerging at $q(a)<2$, the tearing mode cannot be suppressed, but only kept controlled at the edge. At $q(a)>2.5$ or at high density this is not sufficient to prevent a disruption whenever its amplitude grows, even in the absence of a wall-locking. Instead, a robust control is possible when the $m=2$, $n=1$ tearing mode is converted into the non-resonant RWM by dynamically drive the plasma to $q(a)<2$. Even in the presence of a confinement loss produced by the tearing mode, the removal of the magnetic island when $q(a)<2$ allows a full recovery of the discharge. This result has been achieved by applying a pulse on the toroidal voltage through fast solid state power supplies, therefore suddenly increasing plasma current and lowering $q(a)$ below 2, as soon as a $m=2$, $n=1$ tearing exceeding a given amplitude threshold was detected. The control has a 100% success rate even at high densities near to the Greenwald limit. The question is whether this technique could be applied to large conventional tokamaks and possibly extrapolated to ITER. The first requisite is a magnetic feedback control system against the $m=2$, $n=1$ RWM such to make stable the $q(a)<2$ equilibrium. Apart from RFX-mod, this operation has been performed only in DIII-D [35]. The second requisite is the possibility to access rapidly to the $q(a)<2$ configuration, from a standard $q(a)>2$ equilibrium. In RFX-mod this can be done acting on the plasma current, since the toroidal voltage circuit is designed for the demanding high-current RFP operation. On the contrary, this method seems rather unfeasible with the present equipment of conventional tokamaks. Moreover, increasing the current near the machine limits in critical plasma conditions would not be a safe operation. Nonetheless, there is another possible way of changing $q(a)$, particularly attractive for shaped devices. In fact, in a non-circular tokamak the following relationship between the edge $q$ and the elongation $\kappa$ can be established, $q_{edge} = 2\pi a^2 \kappa B_{0\phi}/(\mu_0 I_p R_0)$ [36], being $a$ the plasma half-width. According to this formula passing from an elliptical plasma ($\kappa>1$) to a circular one ($\kappa=1$) will decrease $q_{edge}$ by a factor $\kappa$. Therefore the shape control could realize the transition from a conventional D-shaped equilibrium to a $q(a)<2$ circular configuration. The cross section reduction implied by this transition could increase the plasma resistance, and the toroidal loop voltage might be adjusted to maintain $I_p$ constant. The passage to circular plasma adds two benefits. First, it facilitates the control of the $m=2$, $n=1$ mode, both in the tearing (when $q(a)>2$) and RWM (when $q(a)<2$) typologies, since its spatial structure simplifies through the elimination of the additional poloidal harmonics produced by the D-shape. Second, the elongation ramp-down reduces the confinement and the thermal energy content [37], which, even in the perspective of recovery the discharge, is a prudential action when approaching disruptive conditions. This operation increases the plasma-shell distance, so the destabilization of the $n=1$ ideal external kink mode might become an issue [7]. However, estimates of the critical plasma-shell proximity for the $n=1$ toroidal pressure driven mode give $(r_w/a)_c \approx 1.5$ [38]. Likewise, exploiting the modified Newcomb's criterion [7], we get $(r_w/a)_c \approx 1.2$ for the cylindrical current driven $m=2$, $n=1$. Another cylindrical estimate with the ETAW code [39] provides a more optimistic prediction: $(r_w/a)_c \approx 1.4$. Therefore, there is room to attempt the experiment. Shape control by poloidal field coils in tokamaks is a well-assessed technique [40]. For example, a significant increase of the elongation can be obtained in about 100ms in DIII-D [37], so the opposite transition to circularity should require a similar time-scale. However, only the experiments can tell us if this is fast enough. Perhaps, the competition between the elongation ramp-down time and the current-quench time consequent to a thermal loss, which for JET is of the order of tens of milliseconds [1], it is too unfavorable. However, the elongation ramp-down could, in principle, be activated well in advance to the thermal quench, exploiting, if present, the precursor phase, which can last more than 100ms at JET [2, 41]. The hope is that the transition to the *q(a)*<2 configuration could anticipate the thermal quench, thereby avoiding it, or at least that the restoration of an acceptable confinement consequent to the magnetic island removal could prevent the current quench. In ITER the technological constraints seem more critical. Taking the project discharge time of about 15s for the superconducting poloidal field coils in case of a fault [42], it is clear that any change of the plasma shape will take quite a long time. In any case the data obtained from present elongated tokamak experiments will be extremely important to understand the scalability of this technique to larger and slower devices.

We would like also to emphasize an important point that emerges from our experiments and that somehow conflicts with conventional wisdom in tokamak research. This regards the temperature drop concomitant to the spike in the toroidal loop voltage (and current) characteristic of the disruption process. As shown in figure 9, the thermal quench does not involve a temperature collapse to the few eV level as reported for most tokamaks [2]. We speculate that the sudden impurity contamination, believed to be the ultimate temperature killer [33, 41], does not take place in RFX-mod. Future analyses will try to clarify this issue and to assess whether the feedback control on the tearing mode could have some part in the mitigation of this aspect of the disruption.

A final comment is in order regarding the fact that our experiments have shown that disruptions caused by the *m*=2, *n*=1 tearing mode do not necessarily imply the wall-locking phenomenon. In fact, according to the statistics shown in figure 5 for *q(a)*>2.5 the rotation frequency of the disruptive mode is only one order of magnitude smaller than the natural one. The critical parameters triggering the disruption are the island width and the plasma density. This observation can have implications for ITER, since low frequency mode rotations during disruptions are deleterious when in resonance with the mechanical structure. This is the case for the rotating halo currents during vertical displacement events, as observed in JET [43] and NSTX [44]. Although quantitative predictions for ITER are difficult, the simulation of the tearing mode slowing down is worth to be addressed taking into account the proper electromagnetic boundary of the machine.

**Acknowledgment**

This project has received funding from the European Union Horizon 2020 research and innovation program under grant agreement number 633053. The views and opinions expressed herein do not necessarily reflect those of the European Commision.

**References**

[1] ITER Physics Basis, *Nucl. Fusion* **47** (2007) S128-S202

[2] ITER Physics Basis, *Nucl. Fusion* **39** (1999) 2137

[3] J. A. Wesson, *'Tokamaks'*, Third Edition, Claredon Press-Oxford (2004)

[4] F. Troyon *et al*, *Plasma Phys. Control. Fusion* **26** (1984) 209

[5] M. Greenwald, *Plasma Phys. Control. Fusion* **44** (2002) R27.

[6] A. Bondeson, M. Persson, *Nucl. Fusion* **28** (1988) 1887

[7] R. Fitzpatrick, A. Y. Aydemir *Nucl. Fusion* **36** (1996) 11

[8] M. S. Chu, M. Okabayashi, *Plasma Phys. Control. Fusion* **52** (2010) 123001

[9] P. Martin, *et al, Nucl. Fusion* **51** (2011) 094023

[10] L. Marrelli *et al*, *Proc. 38$^{th}$ Conference on Plasma Physics*, Strasbourg (2011), P. 2091, European Physical Society

[11] P. Zanca *et al*, *Plasma Phys. Control. Fusion* **54** (2012) 094004.

[12] M. Baruzzo *et al*, *Nucl. Fusion* **52** (2012) 103001

[13] R. Paccagnella, 'Tokamak operation at low q and scaling towards a fusion machine', http://arxiv.org/abs/1206.3083

[14] P. Zanca, L. Marrelli, G. Manduchi, G. Marchiori, *Nucl. Fusion* **47** (2007) 1425

[15] M. Okabayashi *et al*, EX/P2-42, 25$^{th}$ IAEA Fusion Energy Conference, Saint Petersburg, Russia (2014).

[16] A. W. Morris, *et al*, *Phys. Rev. Lett* **64** (1990) 1254

[17] G. A. Navratil *et al*, *Phys. Plasmas* **5** (1998) 1855.

[18] L. Novello *et al*, *Fus. Eng. Des.* **86** (2011) 1393-1397

[19] O. Barana, A. Lucchetta, G. Manduchi, C. Taliercio, *Fus. Eng. Des.* **71** (2004) 35

[20] G. Manduchi *et al*, *Fus. Eng. Des.* **87** (2012) 1907

[21] P. Sonato *et al*, *Fus. Eng. Des.* **161** (2003) 66-68

[22] G. Serianni *et al*, *Rev. Sci. Instrum.* **75** (2004) 4338

[23] R. Fitzpatrick, *Nucl. Fusion* **33** (1993) 1049

[24] R. Fitzpatrick, *et al*, *Phys. Plasmas* **6** (1999) 3878

[25] P. Zanca, *Plasma Phys. Control. Fusion* **51** (2009) 015006

[26] P. Zanca, G. Marchiori, L. Marrelli, L: Piron and the RFX-mod team, *Plasma Phys. Control. Fusion* **54** (2012) 124018.

[27] P. Zanca, *Plasma Phys. Control. Fusion* **52** (2010) 115002

[28] R. B. White, D. A. Monticello, M. N. Rosenbluth, B. V. Waddell, *Phys. Fluids* **20** (1977) 800

[29] P. Zanca, D. Terranova, *Plasma Phys. Control. Fusion* **46** (2004) 1115.

[30] P. Innocente, S. Martini, A. Canton, L. Tasinato, *Rev. Sci. Instrum.* **68** (1997) 694

[31] F. Salzedas *et al*, *Phys. Rev. Lett* **88** (2002) 075002

[32] P. Franz *et al*, *Nucl. Fusion* **41** (2001) 695

[33] J. A. Wesson, D. J. Ward, M. N. Rosenbluth, *Nucl. Fusion* **30** (1990) 1011

[34] A. Alfier, R. Pasqualotto, *Rev. Sci. Instrum.* **78** (2007) 013505

[35] P. Piovesan, *et al*, *Phys. Rev. Lett* **113** (2014) 045003

[36] J. Freidberg, *'Plasma Physics and Fusion Energy'*, Cambridge University Press (2007).

[37] L. L. Lao, *et al*, *Phys. Rev. Lett* **70** (1993) 3435

[38] A. Bondeson, D. J. Ward, *Phys. Rev. Lett* **72** (1994) 2709

[39] R. Paccagnella, *Nucl. Fusion* **31** (1991) 1899

[40] G. Ambosino, R. Albanese, *IEEE Control Systems Magazine*, **25** (2005) 76

[41] J. A. Wesson *et al*, *Nucl. Fusion* **29** (1989) 641

[42] A. Winkler, *'Transient behaviour of ITER poloidal field coils'*, KIT Scientific Publishing Karlshrue (2010).

[43] S. Gerasimov *et al.*, *Nucl. Fusion* **54** (2014) 073009

[44] S. Gerhardt, *Nucl. Fusion* **53** (2013) 023005

[45] P. Innocente *et al*, *Nucl. Fusion* **54** (2014) 122001.

[46] C. G. Gimblett, *Nucl. Fusion* **26** (1986) 617.

[47] G. Marchiori *et al*, 'Advances in MHD mode control in RFX-mod', 35[th] EPS conference Hersonissos, Crete, Greece (2008).

[48] L. Carraro et al, *Plasma Phys. Control. Fusion* **40** (1998) 1021

[49] A. Murari, *et al*, *Rev. Sci. Instrum.* **70** (1999) 581

**Appendix A: the RFXlocking model applied to the tokamak**

The cylindrical zero-pressure RFXlocking code, developed for modeling feedback on the RFP dynamo tearing modes [25, 27], has been adapted to study the RFX-mod tokamak discharges. Here we recall the main aspects of the model for the sake of clarity. For a more detailed description we refer to the paper [27]. This is a basic approach describing a classical tearing mode interacting with external conductive structures. An important element is a realistic description of the feedback coils and the control algorithm. The main difference with respect to previous RFP formulations is the inclusion of the Rutherford equation for the mode amplitude evolution at the resonant surface [28]. This equation is in fact a standard model for tokamak tearing modes, but it is of uncertain applicability in the RFP, due to its strong non-linear dynamic. In that case the amplitude at the resonant surface was imposed. We describe plasma equilibrium as a force-free circular cylindrical configuration, with a periodicity length $2\pi R_0$ in the *z* direction, being $R_0$ the plasma major radius: using a right-handed co-ordinate system with a simulated toroidal angle *(r, θ, ϕ≡ z/R_0)* the force balance condition for the equilibrium field $\mathbf{B}_0(r) = (0, B_{0\theta}(r), B_{0\phi}(r))$ is $\nabla \times \mathbf{B}_0 = \mu_0 \mathbf{J}_0 = \sigma(r)\mathbf{B}_0$. As explained in section 3 a two parameters expression for *σ(r)* is adopted. The plasma is contained by a vacuum vessel, modeled as a thin shell with time constant $\tau_v$ placed at *r=r_v*, surrounded by a finite-thickness, time constant $\tau_w$ shell with internal, external radii respectively $r_{wi}$, $r_{we}=r_{wi}+\delta_w$. A regular grid of rectangular active coils is placed outside the shell at *r=c* on a support structure modeled as a thin shell with time constant $\tau_s$. Radial field sensors are considered at some radius $r_s$ between the shell inner surface and the plasma ($a \leq r_s \leq r_{wi}$). For a given equilibrium, and a single *m, n* tearing mode (non-linear coupling is discarded) the zero-pressure Newcomb's equation determines the radial profile of the complex harmonic $\psi^{m,n}(r,t) \equiv -i\, r\, b_r^{m,n}(r,t)$ in the plasma and vacuum regions. At the resonant radius $q(r_{m,n})=m/n$, where it becomes singular, Newcomb's solution is matched with other two equations evolving mode amplitude and phase, defined by $\psi^{m,n}(r_{m,n},t) = \Psi^{m,n}(t) e^{i\varphi^{m,n}(t)}$. Mode amplitude is related to the island width $W$ by the standard formula $W = 4\left( \left.\dfrac{R_0 q^2}{m B_{0\phi}\, r\, dq/dr}\right|_{r_{m,n}} \Psi^{m,n} \right)^{1/2}$, and the latter is evolved by Rutherford equation [23, 28]

A1) $\quad 0.8227 \dfrac{\tau_R}{r_{m,n}} \dfrac{dW}{dt} = r_{m,n} \operatorname{Re}\left[ \dfrac{1}{\Psi^{m,n}} \left( \dfrac{\partial \psi^{m,n}}{\partial r} \right)_{r_{m,n}^-}^{r_{m,n}^+} \right]$

$r_{m,n}^+ = \min(r_{m,n} + W/2,\, a), \quad r_{m,n}^- = \max(r_{m,n} - W/2,\, 0)$

The resistive time in the l.h.s is defined through the Spitzer resistivity $\eta_S$ evaluated at the resonant radius: $\tau_R = \mu_0\, r_{m,n}^2 / \eta_S(r_{m,n})$. The radial derivative at the island boundaries in the r.h.s include the modification to the $\psi$ radial profile produced by the feedback coils and the currents induced on the passive conductive structures.

Mode phase is evolved by [45]

A2) $\quad \dfrac{d\varphi^{m,n}}{dt} = n\, \Omega_\phi(r_{m,n},t) - m\, \Omega_\theta(r_{m,n},t) + \dfrac{nB_{0\theta}}{e n_e R_0 B^2}\left( 1 + \dfrac{m^2}{n^2} \dfrac{R_0^2}{r^2} \right) \left.\dfrac{d(p_e + p_i)}{dr}\right|_{r_{m,n}}$

In this formula the single-fluid (i.e. ion) toroidal and poloidal angular velocities $\Omega_\phi(r,t)$, $\Omega_\theta(r,t)$, averaged over angular co-ordinates, are taken at the resonant radius. The latter term, where $e>0$ is the electron charge magnitude and $n_e$ the electron density ($=n_i$), encapsulates the electron and ion diamagnetic frequencies, which allow passing from the ion to the electron fluid velocity. In fact, above equation states that the island is frozen within the electron fluid, an assumption which gives natural rotation frequencies compatible with the RFX-mod experimental observations both for the tokamak and the RFP configuration (for the latter case see [45]). The single-fluid velocities are evolved by the motion equations, which include the plasma perpendicular viscosity $\mu$ and the electromagnetic torques $\delta T_{EM}$ developed at the resonant radius by the interaction between the island and the external conductive structure:

A3) $\rho \dfrac{\partial \Omega_\phi}{\partial t} = \dfrac{1}{r}\dfrac{\partial}{\partial r}\left(\mu r \dfrac{\partial}{\partial r}\Omega_\phi\right) + S_\phi + \dfrac{\delta T_{EM,\phi}^{m,n}}{4\pi^2 r R_0^3}\delta(r - r_{m,n})$

A4) $\rho \dfrac{\partial \Omega_\theta}{\partial t} = \dfrac{1}{r^3}\dfrac{\partial}{\partial r}\left(\mu r^3 \dfrac{\partial}{\partial r}\Omega_\theta\right) - \dfrac{\rho}{\tau_D}\Omega_\theta + S_\theta + \dfrac{\delta T_{EM,\theta}^{m,n}}{4\pi^2 r^3 R_0}\delta(r - r_{m,n})$

Here $\rho$ is the mass density. The poloidal flow damping operator, expressed with a characteristic time $\tau_D$, and the momentum source densities $S_\phi$, $S_\theta$ are phenomenological terms. For the sake of simplicity we consider all these quantities to be constant with $r$. The electromagnetic torque is expressed by a non-linear combination of the Newcomb's solution taken at the resonant surface and at the plasma-facing conductive structure, i.e. the vacuum-vessel: $\delta T_{EM}^{m,n} \propto \mathrm{Im}\left[\psi^{m,n}(r_v,t)\,\psi^{m,n}(r_{m,n},t)^*\right]$. The imposed velocity boundary conditions are $\partial\Omega_\phi(0,t)/\partial r = \partial\Omega_\theta(0,t)/\partial r = \Omega_\phi(a,t) = \Omega_\theta(a,t) = 0$. The electromagnetic torque modifies the natural plasma velocity, opposed in this action by the viscous torque. The toroidal $\Omega_{\phi 0}(r)$ and poloidal $\Omega_{\theta 0}(r)$ natural angular velocities are derived from (A3), (A4) taking the steady-state ($\partial/\partial t=0$) and discarding the electromagnetic torques:

A5) $\Omega_{\phi 0}(r) = S_\phi \tau_V /(4\rho)\left(1 - r^2/a^2\right),$

$$\Omega_{\theta 0}(r) = S_\theta \tau_D / \rho \left( 1 - a I_1\left(\sqrt{\frac{\tau_V}{\tau_D}} \frac{r}{a}\right) \left( r I_1\left(\sqrt{\frac{\tau_V}{\tau_D}}\right) \right)^{-1} \right)$$

Here $\tau_V = \rho a^2 / \mu$ is the viscous diffusion time an $I_1$ is the modified Bessel function. The Newcomb's solution is also matched with other equations, which model the vessel, shell and coils regions. The thin-shell dispersion relation [46] suitable for the vessel is

A6) $\quad \tau_v \dfrac{\partial \psi^{m,n}}{\partial t}\bigg|_{r_v} = r \dfrac{\partial \psi^{m,n}}{\partial r}\bigg|_{r_v -}^{r_v +}$

Newcomb's solution defines the r.h.s of this equation. The radial variation of the perturbation inside the structure is discarded by (A6). This effect is instead modeled in the shell region according to the more general diffusion equation [46]:

A7) $\quad \mu_0 \sigma_w \dfrac{\partial \psi^{m,n}}{\partial t} = \dfrac{\partial^2 \psi^{m,n}}{\partial r^2}, \quad r \in [r_{wi}, r_{we}]$

Here $\sigma_w$ is the shell conductivity. The radial derivative of the solution of (A7) is matched with Newcomb's solution at $r_{wi}$ and $r_{we}$. Ampere's law applied to the coils region, considered of negligible radial thickness, and a thin-shell relation similar to (A6), which models the structure supporting the coils, provide

A8) $\quad r \dfrac{\partial \psi^{m,n}}{\partial r}\bigg|_{c-}^{c+} = i\mu_0 \left[ m^2 + (nc/R_0)^2 \right] \Im_c(m,n) I_c^{m,n} + \tau_s \dfrac{\partial \psi^{m,n}}{\partial t}\bigg|_c$

In this expression $I_c^{m,n}$ is the coils' current harmonic, and $\Im_c(m,n)$ is the coils' shape factor for the *m, n* mode [27]. An *R/L* modal equation can be cast for the current harmonic:

A9) $\quad \tau_c^{m,n} \dfrac{dI_c^{m,n}}{dt} + I_c^{m,n} \approx \dfrac{N_t}{R_c} V_c^{m,n}$

Here $V_c^{m,n}$ is the Fourier harmonic of the applied voltages, $N_t$ is the number of turns of each coil and $R_c$ is the coil resistance. The self and mutual inductances of the coils synthesize a sort of '$m$, $n$ inductance' of the overall grid, which determines the characteristic time $\tau_c^{m,n}$. In the r.h.s there is also a small extra term (not written) interpreted as the passive stabilizing action of the coils. The voltage harmonic is determined by the control algorithm and by the feedback model. The analysis [27] has shown that a continuous-time feedback model may lead to too optimistic predictions with respect to a discrete-time formulation closer to a realistic digital feedback. Therefore we rely on a discrete-time feedback model in the present analysis. The feedback is characterized by a cycle period $T_f$, according to which the applied voltage is updated, and by a latency $\Delta t$, taken, for sake of simplicity, a multiple of $T_f$: $\Delta t = \bar{k}\, T_f$. The feedback variable $w_f^{m,n}$ is sampled at the discrete times $\{t_k\}$, where $t_j = t_{j-1} + T_f$, and the zero-order hold discretization of the standard proportional-derivative (PD) control is considered:

A10) $\quad V_c^{m,n}(t_j \le t \le t_{j+1}) = \dfrac{R_c}{N_t}\left[ K_p^{m,n} w_f^{m,n}(t_{j-\bar{k}}) + K_d^{m,n} \dfrac{d}{dt} w_f^{m,n}(t_{j-\bar{k}}) \right], \quad K_p^{m,n}, K_d^{m,n} < 0$

This expression, which models both the piece-wise character of the applied voltage and the latency delay, represents a voltage control. Instead, a current control is implemented in RFX-mod, since the coils amplifiers are equipped with an internal feedback circuit [47]. The ensuing improved dynamic response can be simulated by a non-zero derivative gain $K_d$ in (A10). To take into account the finite bandwidth of the sensors and the filtering required by the digital acquisition we distinguish the acquired feedback variable $w_f^{m,n}$, used by the feedback system, from the 'physical' feedback variable $b_f^{m,n}$. The two can be related by a one-pole filter law

A11) $\tau_d \dfrac{d}{dt} w_f^{m,n}(t) + w_f^{m,n}(t) = b_f^{m,n}$

where the time constant $\tau_d$ is set equal to $T_f$. The system of equations is closed by the definition of $b_f^{m,n}$. Assuming CMC [14], where the impact of the coils' sideband harmonics is minimized, we take the $m, n$ component of the radial field at the sensor radius: $b_f^{m,n} = b_r^{m,n}(r_s)$.

*Parameter settings (in MKS units)*

Here we list the typical values set in simulation for the above defined parameters. The RFX-mod front end is adopted: $a$=0.459, $R_0$=2, $r_v$=0.49 (vessel average radius), $\tau_v$=0.0023 (for the $m$=2, $n$=1 mode), $r_{wi}$=0.5125, $\delta_w$=0.003, $\tau_w = \mu_0 \sigma_w r_{wi} \delta_w$=0.1 (copper shell), $c$=0.5815, $\tau_S$=0.024. On-Axis toroidal field is set to the typical value 0.55T for the RFX-mod tokamak discharges. Particle density is fixed constant with radius at the value $n_e$=$n_i$=0.2$n_G$. The perpendicular viscosity $\mu$ is defined through the viscous diffusion time, taken of the same order of the estimated energy confinement time: $\tau_V$=0.06. Moreover we set $\tau_D/\tau_V$=0.001 to simulate a strong poloidal flow damping. The momentum source $S_\phi$ is fixed by the first of (A5), imposing at half plasma radius the toroidal velocity as measured by the passive Doppler spectroscopy based on the OV emission [48]: $V_{0\phi}(a/2) = R_0 \Omega_{0\phi}(a/2) = -1.25 \times 10^4$. Moreover we take $S_\theta$=$S_\phi$. As far as temperature is concerned we assume $T_e$=$T_i$ with a parabolic radial profile reaching 250eV on-axis. The RFX-mod values $R_c$=0.8 and $N_t$=60 in equation (A9), and $T_f$=2×10$^{-4}$, $\Delta t$=6$T_f$ for equation (A10) are imposed. Finally, $\tau_c^{m,n}$ is estimated about 0.003, and we set $K_d^{m,n}$=2$\tau_c^{m,n} K_p^{m,n}$ to mimic the improved transient response of the RFX-mod coils' current control.

**Appendix B: *q(0)* estimate in the RFX-mod tokamak discharges**

Two different approximate methods to determine the on-axis $q$ are described: the first stems from the internal inductance $l_i = 2 \int_0^a r B_{0\theta}^2 dr \big/ \left(a^2 B_{0\theta}^2(a)\right)$ provided by the edge magnetic measurements; the second is a local estimate based on a modified Spitzer's expression for the resistivity and the core plasma temperature measurement. Both rely on cylindrical geometry approximation.

## $q(0)$ from $l_i$ and $q(a)$

Upon assuming the above mentioned large aspect ratio, low-beta force-balance condition $\mu_0 \mathbf{J}_0 = \sigma(r)\mathbf{B}_0$, $\sigma(r) = \sigma_0 \cdot \left(1-(r/a)^2\right)^\alpha$, the connection between $q(0)$, $q(a)$ and $l_i$ is established by the two approximate relations [3]:

B1) $\quad l_i \approx \ln(1.65 + 0.89\alpha); \quad \alpha \approx q(a)/q(0) - 1$

Formulas (B1) allows computing $q(0)$ from $l_i$ and $q(a)$. The cylindrical $q(a)$ is easily obtained by the edge magnetic measurements, whereas $l_i$ is got from the standard relation $\Lambda = \beta_p + l_i/2 - 1$, where $\Lambda$ is the asymmetry factor of the poloidal field (computed from edge measurements) and $\beta_p$ is the poloidal beta. Given the difficulties of the temperature measurement in the RFX-mod low-current tokamak plasmas, we prefer to evaluate $\beta_p$ from the diamagnetic flux $\Delta\Phi$. In fact, in cylindrical geometry we have $\beta_p = 1 + 8\pi B_{0\phi}(a)/(\mu_0 I_p)^2 \Delta\Phi$ [3], where the cylindrical estimate of $\Delta\Phi$ is:

B2) $\quad \Delta\Phi = \pi a^2 B_{0\phi}(a) - \xi_{tor,cyl} \Phi(a)$

Here, $\Phi(a)$ is the measured toroidal flux enclosed by the plasma and $\xi_{cyl,tor}$ is a geometrical conversion factor from toroidal to cylindrical geometry. This is determined in vacuum shots by the ratio $\xi_{tor,cyl} = \left(\pi a^2 B_{0\phi}(a)/\Phi(a)\right)_{vacuum}$.

## $q(0)$ from $T_e$

From Ohm's law and the previous equilibrium model applied at $r=0$ one gets:

B3) $\quad E_{0\phi}(0) = \eta(0) J_{0\phi}(0); \quad \mu_0 J_{0\phi}(0) = \sigma_0 B_{0\phi}(0); \quad \sigma_0 = 2/(q(0) R_0)$

In stationary conditions $E_{0\phi}$ has a constant radial profile and can be determined by the loop-voltage external measurement: $E_{0\phi}(0) = V_\phi/(2\pi R_0)$. By combining the previous expression and approximating $B_{0\phi}(0) \cong B_{0\phi}(a)$ one gets $q(0) V_\phi/B_\phi(a) \cong 10^7 \times \eta(0)$. We model the central resistivity with the Spitzer's expression plus an offset:

B4) $\quad q(0) \dfrac{V_\phi}{B_\phi(a)} \cong Z_{eff}\, \eta_0\, T_e(0)^{-3/2} + C$

The central electron temperature (in (B4) expressed in keV) is measured by the double filter SXR diagnostic [49]. The two quantities $V_\phi/B_\phi(a)$ and $\eta_0 T_e(0)^{-3/2}$ have been correlated in the flattop stationary phase of some H, D and He discharges: a linear fit with a non-zero offset $C_{fit}>0$ and angular coefficients close to 1 for H, D and 2 for He shots interpolates reasonably well the data. This implies that *q(0)* is about 1 on average, so fixing $Z_{eff}=1$ for the H, D and $Z_{eff}=2$ for He we may interpret the data dispersion around the fit mainly as an effect of the *q(0)* deviation from unity. Therefore, the following estimate can be provided:

B5) $\quad q(0) \cong \left(Z_{eff}\, \eta_0\, T_e(0)^{-3/2} + C_{fit}\right) \dfrac{B_\phi(a)}{V_\phi}$

In figure 14 the two estimates of *q(0)* show a similar trend to increase with *q(a)*. This is qualitatively confirmed by an inspection of the saw-tooth activity on the SXR signal, whose detection implies *q(0)<1*.

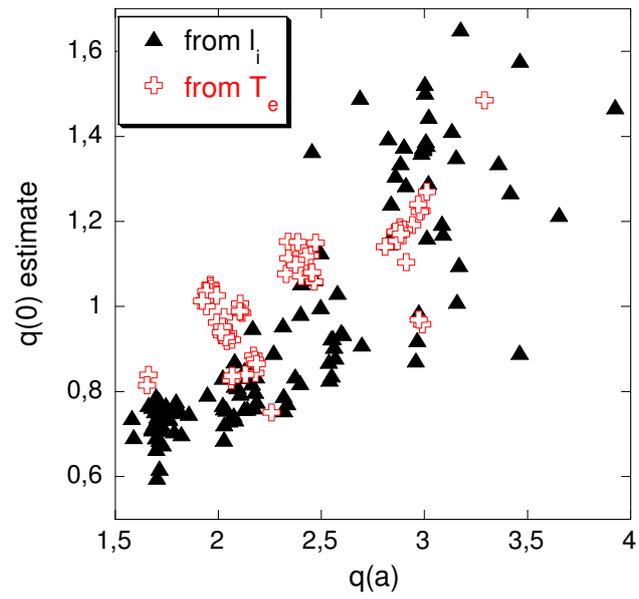

**Figure 14**. Summary plot of the two different *q(0)* estimates.